instructions: there are four postscript files for four figures appended
              after the tex file(below \end{document}) in the order of
              Fig.1,2,3 and 4, each of which begins with "
              with "showpage".

\documentstyle[preprint,aps]{revtex}
\begin{document}
\draft
\title{
Fractional Quantum Hall Effect, Cranked Harmonic Oscillator,\\
and Classical Periodic Orbits
}
\author{R.K.Bhaduri, Shuxi Li, Kaori Tanaka\\
and\\
J.C.Waddington}
\address{
Department of Physics and Astronomy, McMaster University,\\
Hamilton, Ontario, Canada L8S 4M1
}
\maketitle
\begin{abstract}

A two-dimensional harmonic oscillator, when rotated by the oscillator
frequency, generates Landau-like levels.  A further cranking results
in condensates and gaps resembling the fractional quantum Hall effect.
For a filling fraction $\nu=p/q$, with $q$ odd, the model predicts
that the gap is
proportional to $\sqrt{n_{e}/pq}$, where $n_{e}$ is the electron
density of the sample.  This agrees well with recent experimental data.
Qualitative arguments for the success of the model are given.
\end{abstract}

In this paper, it is first recalled that a two-dimensional harmonic
oscillator, when rotated by the oscillator frequency, generates
Landau-like levels.  A new result is then proven : that a
further cranking of the oscillator gives rise to a succession
of quantum gaps and condensates at particular values of the
incremental frequency.  When expressed as dimensionless fractions,
these reduce, for the odd denominators, to the Haldane$^{1}$
hierarchy in the fractional quantum Hall effect (FQHE).
It is of interest to investigate whether or not the gaps in this
model have any relationship to those observed$^{2}$ in the
FQHE.  The phenomenon of FQHE is reasonably well understood
$^{3-5}$, though it continues to draw both experimental$^{6,7}$
and theoretical$^{8-11}$ attention.  It has been established$^{3}$
that in the FQHE, the electrons in the lowest Landau level are
strongly correlated due to the repulsive Coulomb interaction
between the pairs, and the quantum gaps arise from this effect.
It is shown in this paper that the centrifugal term in the
rotating frame gives rise to a repulsive correlation in the
two-particle coordinate similar to the Coulomb potential.
The result of this equivalence is that the gaps must vary as
$\sqrt{B}$, where $B$ is the applied external magnetic field.
The predictions of the cranking model, at a quantum level, are
compared with the latest experimental data.$^{7}$
The close connection between the quantum gaps and the classical
periodic orbits is emphasized.$^{12}$  Finally the implication of the
success of the model is discussed.

The integral quantum Hall effect (IQHE) can be explained in
the independent particle model in conjunction with the
localized states due to impurities.  To obtain the gaps
by the cranked oscillator in this case, consider the single-electron
Hamiltonian $H=\bigl(\,{\bf p}-{e \over c}{\bf A}\,\bigr)^{2}/2M$.
Here the vector potential ${\bf A}={1 \over 2}({\bf B}{\times}{\bf r})$,
$e=-\vert e \vert$, and the motion is confined in the $xy$ plane.
In this symmetric gauge, $H$ reduces to the form
\begin{equation}
H={1 \over 2M}(\,p_{x}^{2}+p_{y}^{2}\,)
+{1 \over 2}M\biggl({{\omega}_{c} \over 2}\biggr)^{2}(x^{2}+y^{2})
+{{\omega}_{c} \over 2}l_{z}\;,
\end{equation}
where ${\omega}_{c}={{\vert e \vert B} \over Mc}$, and
$l_{z}=xp_{y}-yp_{x}$.
Note that Eq.($1$) may be interpreted as the
Hamiltonian of a particle in a harmonic oscillator of Larmor frequency
$\omega_{c}/2$, that is itself rotating about the (negative)
$z$-axis with the same frequency.  The collapse of the single-particle
states into degenerate Landau levels may be shown graphically by
considering an auxiliary Hamiltonian $\tilde H$,
\begin{equation}
{\tilde H}={1 \over 2M}(\,p_{x}^{2}+p_{y}^{2}\,)
+{1 \over 2}M\biggl({{\omega}_{c} \over 2}\biggr)^{2}(x^{2}+y^{2})
+{\omega}l_{z}=H+{\tilde \omega}l_{z}\;,
\end{equation}
where ${\tilde \omega}=(\omega - \omega_{c}/2)$ and $H$ is defined
by Eq.$(1)$.  Fig.$1$ depicts the spectrum of the energy levels
as a function of ${\tilde \omega}/\omega_{c}$ in the range
$-1/2 \le {\tilde \omega}/\omega_{c} \le 1/2$.  The Landau gaps
of $\hbar \omega_{c}$ appear at ${\tilde \omega}/\omega_{c}=0$,
while the uncranked oscillator spectrum of Larmor frequency
$(\omega_{c}/2)$ is at ${\tilde \omega}/\omega_{c}=-1/2$.
Note that for certain values of parameter ${\tilde \omega}/\omega_{c}$,
a sequence of well-defined gaps develop.  The repeating pattern
for ${\tilde \omega}/\omega_{c}\ge 0$ is known as a Farey fan$^{13}$,
and has been studied in the context of number theory and
continued fractions.  Since the gaps that appear in the range
$1/2 < {\tilde \omega}/\omega_{c} \le 1$ are symmetric about
${\tilde \omega}/\omega_{c}=1/2$, the horizontal axis is terminated
at $1/2$.  As Fig.$1$ shows, the angular momentum states
collapsing at the lowest Landau level are all aligned in the same
(negative $z$) direction, and each value of the angular momentum
shows up only once.  At the same energy $E={1\over 2}\hbar\omega_{c}$,
inspection of the level at ${\tilde \omega}/\omega_{c}=1/m$
($m$ an integer $>1$) reveals that the number of converging
single-particle states is exactly a fraction $1/m$ of the Landau
levels.  For example, the successive harmonic oscillator states
meeting at ${\tilde \omega}/\omega_{c}=1/3$, with energy
$E={1\over 2}\hbar\omega_{c}$ have angular momenta $0,-3,-6, etc.$
in units of $\hbar$ (see Fig.$2$).  So for every triplet of adjacent
states of the Landau level, $e.g.$, $(0, -1, -2)$, there is one
($l=0$ in this case) at ${\tilde \omega}/\omega_{c}=1/3$.
To construct a stable model, assume the ``sea'' of states
below $E={1\over 2}\hbar\omega_{c}$ to be all occupied, and the states
with $E>{1\over 2}\hbar\omega_{c}$ to be empty.  Then the number of electrons
per unit area occupying the level with ${\tilde \omega}/\omega_{c}=1/3$
is exactly ${1\over 3}n_{1}$, where $n_{1}={{\vert e \vert B_{1}}\over{hc}}$
is
the degeneracy of a Landau level.  Therefore the dimensionless
parameter ${\tilde \omega}/\omega_{c}$ for nonzero positive values
can be identified with the ``filling fraction'' $\nu$ of the FQHE.
This has a curious implication on the statistics of a daughter
state, $e.g.$, at ${\tilde \omega}/\omega_{c}=2/5$.  As shown in
Fig.$2$, one state from each quantum shell of the $\nu=1/3$
mother converges at $E={1\over 2}\hbar\omega_{c}$, constituting
this daughter at $2/5$, just as the mother herself was formed
from the collapse of the states from each Landau level.
Note that the successive states collapsing at $2/5$ have
angular momenta $0, -5, -10, etc.$, so there are only one fifth as
many states as in the Landau level, despite the electron filling
fraction being two fifths.  For this sequence of
daughters, $p\over{2p+1}$, with $p=1,2,3,...^{8}$, the occupancy
is $p$ times the number of states.  Furthermore Fig.$2$
illustrates that the gaps for this same sequence are governed by
the right-angled triangle $abf$.  At the end-point, $\nu=1/2$,
however, the degeneracy of the level again matches the occupancy.

Comparison with the experimental data$^{7}$ can be made
directly by noting, from the triangle $abf$ in Fig.$2$,
that the model gaps $\tilde \Delta$ are proportional to
$(\nu-1/2)$.  Since $\nu={{h c}\over{\vert e \vert B}}n_{e}$
for a sample with electron density $n_{e}$, it follows that
$\tilde \Delta$ should vary linearly when plotted against $1/B$.
This is shown to be the case in Fig.$3$, where the magnitude of the negative
intercept on the $y$-axis (arising from the finite width of the
levels$^{7}$) plus the experimental $\Delta$ may be interpreted
as $\tilde \Delta$.  This plot is to be contrasted with Fig.$3$
of Ref.$7$, where the same gaps were plotted as a linear function of $B$.
Both appear to be valid, since the range of $B$ is limited.
Comparison with experiment over a wider range of a variable
may be made by combining some theoretical input with the model
prediction (as seen in Fig.$1$) that the gap at $\nu=p/q$ varies
as $\hbar \omega_{c}/q$.
It is well-known that the energy scale in FQHE is determined
by the Coulomb term $e^{2}\over {\epsilon l_{0}}$, rather than
$\hbar \omega_{c}$, where
$l_{0}=\sqrt{{\hbar c}\over{\vert e \vert B}}$ is the
magnetic length.  For this reason, in Fig.$2$, the mother-daughter
sequence of condensations are shown with the $y$-axis for
the energy scale in units of $C {e^{2}\over{\epsilon l_{0}}}$,
where $C$ is a dimensionless constant.  This procedure of
replacing the energy unit $\hbar \omega_{c}$ by something
proportional to the Coulomb term will be justified presently
from the dynamical model.  From Fig.$2$, it then follows that
the FQHE gap at $\nu=p/q$ is given by
\begin{equation}
{\tilde \Delta}_{p/q}={1\over q}C{e^{2}\over{\epsilon l_{0}}}=
C{e^{2}\over \epsilon}\sqrt{{2\pi n_{e}}\over{pq}}\;,
\end{equation}
where $n_{e}=\nu/(2 \pi l_{0}^{2})$.  That the gap is proportional to
the quantity $\sqrt{n_{e}\over{pq}}$, is another prediction of
the model.  This is tested in Fig.$4$, where the experimental
data are seen to obey this relation over a
wide range of $\sqrt{1\over{pq}}$.
In order to fit the magnitudes of the gaps by Eq.$(3)$,
it is found that $C \simeq 0.15$ for $\nu=1/3$.

It is important to remember that in a single-particle Hamiltonian
(like $\tilde H$ of Eq.$(2)$), quantum gaps and classical periodic
orbits are closely linked$^{12}$.  The classical equations
of motion of a particle are easily obtained from $\tilde H$,
and may be expressed in a compact form in terms of the
variable $z=x+i\,y$ :
\begin{equation}
\ddot {z}=2i\omega\dot {z}+
\bigl(\,\omega^{2}-{\omega_{c}^{2} \over 4}\,\bigr)z\;.
\end{equation}
For the Landau orbits, $\omega=\omega_{c}/2$, so the centrifugal
term drops out, giving $\ddot {z}=i\omega_{c}\dot {z}$, and only
circular solutions are obtained.
The general solution of Eq.($4$) is given by
$z=Ae^{i(\omega-\omega_{c}/2)t}+Be^{i(\omega+\omega_{c}/2)t}$.
When the ratio of the normal modes $\tilde \omega$ and
$(\tilde \omega + \omega_{c})$ is a rational fraction,
the resulting orbit in the $xy$ plane is periodic.
The first generation mothers in the present model have the
normal modes in the ratio $(\tilde \omega + \omega_{c})/\omega=(m+1)$,
yielding $\tilde \omega/\omega_{c}=1/m$, matching with the
occurrence of the quantum gaps.  More generally, if
$\tilde \omega/\omega_{c}$ equals the irreducible fraction $p/q$,
a classical periodic orbit of $q$-fold symmetry is obtained.
The corresponding quantum gap, from Fig.$1$, is seen to be
${\hbar \omega_{c}}\over q$.

It is now appropriate to discuss the rationale behind the cranking
model.  To see how the Coulomb repulsion between two electrons
may be simulated by the centrifugal term, consider two
noninteracting particles, each with an effective mass $M^{*}$,
being cranked about the $z$-axis as in Eq.$(2)$.  The two-particle
Hamiltonian may be easily separated into the relative
and the centre-of-mass coordinates.  The cranking term is still
${\tilde \omega} l_{z}$, but now $l_{z}$ is the relative angular
momentum ${1\over 2}({\bf r}_{1}-{\bf r}_{2})\times
({\bf p}_{1}-{\bf p}_{2})$ between the two particles.
For a mother $\nu=1/m$, it was noted that the adjacent
collapsing levels differ in their angular momenta by the
magnitude $m \hbar$.  For such a pair, the additional term
${\tilde \omega} l_{z}=({{\tilde \omega}\over \omega_{c}})\omega_{c}
l_{z}=\hbar \omega_{c}$.  For this term to simulate the repulsive
Coulomb repulsion, it follows that $\hbar \omega_{c}=C
{e^{2}\over {\epsilon l_{0}}}$, as was done in Fig.$2$.
This, of course, is only possible if $\omega_{c}$ is replaced
by $\omega_{c}^{*}={{\vert e \vert B}\over{M^{*} c}}$,
and $M^{*} \propto \sqrt{B}$.  Furthermore, for the classical motion
in relative coordinates, an equation analogous to Eq.$(4)$ is
obtained, where $z$ now refers to the relative coordinates.
Note that the effective centrifugal term in Eq.$(4)$ is
repulsive for $\omega > \omega_{c}/2$, and is proportional
to $z$.  On the other hand, the Coulomb potential between
two electrons is $e^{2}\over{\epsilon (x^{2}+y^{2})^{1/2}}$,
and the resulting repulsive force is ${e^{2} z}\over{\epsilon
(z z^{*})^{3/2}}$.  Since the relative distance
$(x^{2}+y^{2})^{1/2}$ is determined by the density of the
electrons (which in turn is fixed by the magnetic length),
it follows that the Coulomb repulsion between two electrons also
varies linearly with $z$.  Thus the cranking model generates an
effective repulsion between adjacent particles that may simulate
the Coulomb repulsion by choosing the energy scale suitably.

The many-body wave function in the cranking model at $\nu=1/m$
may be easily calculated using the harmonic oscillator states.
It does not, of course, yield the form of the correlated
Laughlin wave function.  Apparently, the cranking model condensate
for a mother (or a daughter) contains components from every
Landau level, which is totally different from the Laughlin
state.  This may not really be the case since the energy scale
$\hbar \omega_{c}$ is being replaced by the Coulomb term
$C {e^{2}\over{\epsilon l_{0}}}$ in the cranking model.
Nevertheless, it appears that the full complement of the
Laughlin correlations is not necessary to obtain the FQHE
gaps.  The situation is reminiscent of the nuclear
(incompressible) fluid in an atomic nucleus, where the Jastrow$^{14}$
correlations are not needed to obtain the shell gaps at the magic
numbers.  Indeed, the superdeformed nuclei$^{15}$ discovered
recently have quantum gaps$^{16-17}$ related to classical
periodic orbits similar to the model presented here.

\acknowledgments

This research was supported by grants from NSERC (Canada).
The authors are very grateful to Professors Tsuneya Ando, Matthias Brack
and Akira Suzuki and the colleagues at McMaster for discussions.
Ranjan Bhaduri is thanked for giving Ref.$13$.

\begin{figure}
\caption{The energy spectrum (Eq.($2$)) of a cranked two-dimensional
harmonic oscillator, shown as a function ${\tilde \omega}/\omega_{c}$,
where ${\tilde \omega}=(\omega-\omega_{c}/2)$.  The angular momenta
$l$ of the first few states collapsing at the lowest Landau level
are also shown.  Note the appearance of gaps at particular values
of ${\tilde \omega}/\omega_{c}$.  The pattern (in the range
$1/2 < {\tilde \omega}/\omega_{c} \le 1$ is symmetrical about
${\tilde \omega}/\omega_{c}=1/2$, and is not shown.}
\end{figure}
\begin{figure}
\caption{The mother-daughter sequence of condensed states.
The energy scale in the $y$-axis has an overall scaling factor
$C$ (see text). One
particle state from each Landau level, of alternating parity,
condense at ${\tilde \omega}/\omega_{c}=1/3$ to form a mother.
Similarly, one state from each shell at $1/3$ condense at $2/5.$
Note that a similar condensation takes place from the holes at
$1/3$, to form a hole-daughter at $2/7$.  Only a few converging
lines are shown for clarity.  In the triangle $abf$, the vertical
lines $ab$ and $cd$ show the gaps at $\nu=1/3$ and $2/5$
respectively.}
\end{figure}
\begin{figure}
\caption{The FQHE experimental gaps $\Delta$ (in degrees $K$),
taken from Fig.$3$ of Ref.$7$, are plotted as a function of
$1/B$, where $B$ is the applied external field in Tesla.
The data refer to two different samples (squares and diamonds
: $n_{e}=1.12\times 10^{11}/{\rm cm}^{2}$, triangles and inverted
triangles : $n_{e}=2.3\times 10^{11}/{\rm cm}^{2}$).  The diamonds
and inverted triangles have mother $1/3$, squares and triangles
refer to mother $2/3$ and her daughters.
The two crosses belong to a different sequence with mother $1/5$.}
\end{figure}
\begin{figure}
\caption{The experimental gaps$^{7}$, divided by $\protect\sqrt{n_{e}}$,
are plotted in arbitrary units as a function of $1/\protect\sqrt{pq}$,
where $\nu=p/q$ is
an irreducible fraction.  All four branches of the data points of
Fig.$3$ fall on a single straight line.  The notation is the same
as in Fig.$3$.
}
\end{figure}
\end{document}